\documentclass[fleqn,10pt]{wlscirep}
\usepackage[utf8]{inputenc}
\usepackage[T1]{fontenc}
\usepackage{dcolumn}
\usepackage{bm}
\usepackage{bbm}
\usepackage{lineno}

\usepackage{color}
\definecolor{qasm_red}{RGB}{214, 39, 40}
\definecolor{qasm_blue}{RGB}{0, 0, 255}
\definecolor{hanoi_red}{RGB}{139, 0, 0}
\definecolor{hanoi_blue}{RGB}{0, 0, 139}

\newcommand{\qasmblue}[1]{\textcolor{qasm_blue}{#1}} 
\newcommand{\qasmred}[1]{\textcolor{qasm_red}{#1}} 
\newcommand{\hanoiblue}[1]{\textcolor{hanoi_blue}{#1}} 
\newcommand{\hanoired}[1]{\textcolor{hanoi_red}{#1}} 

\newcommand{\hanoiB}{\ding{53}}
\newcommand{\hanoiN}{\ding{59}}
\newcommand{\qasm}{\ding{108}}

\usepackage{graphicx}
\usepackage{dcolumn}
\usepackage{bm}
\usepackage{pifont}

\title{Resource-efficient quantum correlation measurements via multicopy
neural network methods}

\author[1,2,*]{Patrycja Tulewicz}
\author[1,3,+]{Karol Bartkiewicz}
\author[1,3,$\dagger$]{Adam Miranowicz}
\author[3,4,5,$\ddagger$]{Franco Nori}
\affil[1]{Institute of Spintronics and Quantum Information, Faculty of Physics and Astronomy, Adam Mickiewicz University, PL-61-614 Pozna\'n, Poland}
\affil[2]{QSLab: Quantum Software Laboratory, Department of Engineering and Architecture (DIA), University of Parma, Parma, 43124, Italy}
\affil[3]{Theoretical Quantum Physics Laboratory, Cluster for Pioneering Research, RIKEN, Wakoshi, Saitama 351-0198, Japan}
\affil[4]{Center for Quantum Computing, RIKEN, Wakoshi, Saitama 351-0198, Japan}
\affil[5]{Department of Physics, University of Michigan, Ann Arbor, Michigan 48109-1040, USA}

\affil[*]{patrycja.tulewicz@amu.edu.pl}

\affil[+]{karol.bartkiewicz@amu.edu.pl}
\affil[$\dagger$]{adam.miranowicz@amu.edu.pl}
\affil[$\ddagger$]{fnori@riken.jp}

\begin{abstract}
Measuring complex properties in quantum
systems, such as measures of quantum entanglement and Bell
nonlocality, is inherently challenging. Traditional methods, like
quantum state tomography (QST), require a full reconstruction of
the density matrix for a given system and demand resources that
scale exponentially with system size. We propose an alternative
strategy that reduces the required information by combining
multicopy measurements with artificial neural networks (ANNs),
resulting in a 67\% reduction in measurement requirements compared to
QST. We have successfully measured two-qubit quantum correlations
of Bell states subjected to a depolarizing channel (resulting in
Werner states) and an amplitude-damping channel (leading to
Horodecki states) using the multicopy approach on real quantum
hardware. Our experiments, conducted with transmon qubits on IBMQ
quantum processors, quantified the violation of Bell's inequality
and the negativity of two-qubit entangled states. We compared
these results with those obtained from the standard QST approach
and applied a maximum likelihood method to mitigate noise. We
trained ANNs to estimate degrees of entanglement and nonlocality measures
using optimized sets of projections identified through Shapley's
(SHAP) analysis for the Werner and Horodecki states. The ANN
output, based on this reduced set of projections, aligns well with
expected values and enhances noise robustness. This approach
simplifies and improves the error robustness of multicopy
measurements, eliminating the need for complex nonlinear equation
analysis. It represents a significant advancement in AI-assisted
quantum measurements, making the practical implementation on current
quantum hardware more feasible.
The experimental results demonstrate improved noise robustness on the current noisy 
intermediate-scale quantum (NISQ) hardware, representing a practical advance in 
resource-efficient characterization of quantum correlations.
\end{abstract}
\begin{document}

\flushbottom
\maketitle

\thispagestyle{empty}

\section*{Introduction}

Quantum entanglement~\cite{Schrodinger1935,Einstein1935,RevModPhys.81.865} 
and Bell's nonlocality, here referred to as the degree of Clauser-Horne-Shimony-
Holt (CHSH) inequality violation~\cite{Bell1964,bell_aspect_2004,RevModPhys.86.419}, 
are the bedrocks of quantum engineering, quantum information 
processing~\cite{nielsen_chuang_2010}, quantum communications (e.g., quantum 
teleportation) \cite{PhysRevLett.92.047904}, and quantum cryptography 
(e.g., quantum key distribution)~\cite{PhysRevLett.67.661}.
Quantum nonlocality has become the foundation for numerous information processing 
protocols implemented within quantum systems~\cite{PhysRevA.56.1201,
Mayers1998QuantumCW,PhysRevLett.98.230501}. Measuring entanglement and Bell's 
nonlocality helps to validate various predictions of quantum mechanics and ensures 
that quantum systems behave as expected. At the same time, this pursuit is pushing 
the boundaries of our experimental capabilities, driving advancements in quantum 
technologies.
The efficient and accurate measurement of quantum correlations is not merely a theoretical pursuit but has significant practical implications for quantum 
technology development. Entanglement verification serves as a critical diagnostic 
tool for validating quantum hardware, characterizing quantum channels, and 
assessing the performance of quantum communication protocols. As quantum devices 
scale beyond a few qubits, traditional approaches become prohibitively expensive 
in terms of measurement resources, making efficient alternatives increasingly 
valuable for practical applications.
However, measuring these properties has traditionally relied on
quantum state tomography (QST), which becomes impractical for large-scale qubit systems because
the required QST measurements scale exponentially. Several alternative detection methods for entanglement 
and nonlocality have been proposed in the literature. These include adaptive 
approaches that optimize measurement settings based on partial 
results~\cite{PhysRevLett.105.230404}, collective witness methods utilizing 
multiple copies of a quantum system~\cite{PhysRevLett.89.127902,
PhysRevLett.90.167901,PhysRevLett.95.240407,PhysRevLett.107.150502,
PhysRevA.94.052334}, randomized measurement techniques~\cite{elben2020cross}, 
and entanglement witnesses designed to be robust against 
noise~\cite{PhysRevLett.118.150505}. Particularly notable are 
measurement-device-independent (MDI) approaches~\cite{Shahandeh2016} that allow 
entanglement quantification without trusting the measurement apparatus through 
semi-quantum nonlocal games, offering inherent resilience against certain 
measurement errors. Additionally, machine learning approaches have emerged that 
can extract entanglement information from incomplete measurement 
data~\cite{doi:10.1126/sciadv.add7131,10.1088/1572-9494/ad4090,Pan2024,Kadlec2024}. 
While each method offers advantages in specific scenarios, practical implementation 
on current quantum hardware remains challenging, particularly when balancing 
measurement efficiency, accuracy, and noise resilience.

\begin{figure*}[htb!]
\includegraphics[width=1\linewidth]{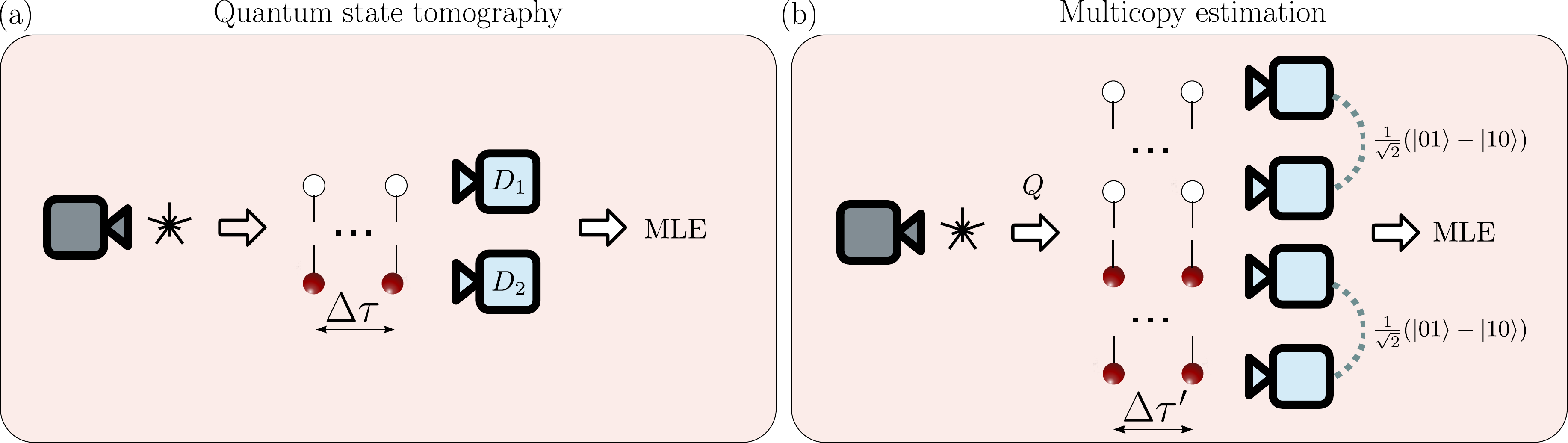}
\caption{\label{fig:idea} Schematic representation of the approaches based on (a) quantum state 
tomography (QST) and (b) multicopy estimation (MCE). 
The symbols $\Delta \tau$ and $\Delta \tau'$ denote the delay time between the generation of each entangled pair
or a set of copies of entangled pairs, respectively.
In (a), detectors $D_1$ and $D_2$ measure products of eigenstates of Pauli's matrices. The multicopy state generation uses a quantum memory $Q$ to store and release the collected pairs.
The detection rates are then processed using maximum likelihood estimation (MLE)
to obtain the most probable physical combination of the measured experimental settings.}
\end{figure*}

In this paper, we present a comprehensive approach to overcome these limitations by 
combining a multicopy estimation (MCE) methodology with artificial neural 
networks. Building upon the theoretical foundations of multicopy measurement 
developed in~\cite{PhysRevA.95.022331,PhysRevA.97.012107}, we introduce several 
significant improvements that bridge the gap between theoretical possibilities and 
practical implementation on current quantum hardware.
Our contributions address multiple challenges in quantum correlation measurement 
through a cohesive framework. First, we demonstrate the first experimental implementation of multicopy measurements on currently available quantum processors 
and simulators, directly confronting practical challenges of noise, limited 
connectivity, and finite sampling statistics that have previously hindered the application of these techniques. Second, we introduce a Maximum Likelihood Estimation (MLE) framework specifically tailored for multicopy measurements that 
substantially improves robustness to experimental noise. This novel approach takes 
into account the polynomial nature of higher-degree multicopy observables, 
providing a significant advantage over the quadratic optimization traditionally 
used in QST.
To further enhance efficiency, we apply systematic SHAP (SHapley Additive 
exPlanations) analysis~\cite{lundberg2017unifiedapproachinterpretingmodel} to identify the minimal set of measurements required for reliable entanglement and 
Bell nonlocality estimation. This innovative application of SHAP enables us to 
dramatically reduce the experimental complexity from 15 measurements to just 5, 
representing a 67\% reduction in measurement requirements. We then demonstrate how neural networks can process this measurement data with enhanced noise 
robustness compared to direct computation, creating a practical pathway for 
implementation on noisy intermediate-scale quantum (NISQ) devices.
Our work also includes a comprehensive comparison with randomized measurement 
techniques~\cite{elben2020cross}, providing a detailed analysis of relative gains 
in query complexity, fault tolerance, and resource requirements. Finally, we 
analyze the scalability of our approach for larger quantum systems, demonstrating 
more favorable scaling characteristics for pairwise correlations compared to 
traditional methods.
A key advantage of our approach is the substantial reduction in query complexity 
compared to traditional QST. For a two-qubit system, QST requires 15 different 
measurement settings (3 measurement settings per qubit for each of the 4 
subsystems, plus joint measurements). In contrast, our multicopy method requires at most 11
measurements to determine the entanglement measure (such as
negativity) and 12 measurements to evaluate a Bell-CHSH
nonlocality measure for arbitrary two-qubit states. Most importantly, 
by integrating an artificial neural network (ANN) with our optimized SHAP selection method, we demonstrate that 
only $5$ carefully selected measurements are sufficient for accurate entanglement 
quantification. This represents a $67\%$ reduction in the number of measurements 
required compared to QST, a significant resource savings for experimental 
implementations.
The advantage of our approach becomes even more pronounced as the system size 
increases. Regarding a system with $n$ qubits:
(i) For QST, $O(4^n)$ measurement settings are needed, and the scaling increases exponentially with system size.
(ii) Multicopy with a neural network has a constant factor improvement over QST and scales as $O(2^n)$.
While our method still exhibits exponential scaling with system size, the reduction 
in the exponent's base from 4 to 2 represents a quadratic improvement in scaling 
behavior. This translates to a substantial practical advantage for moderately sized 
systems relevant to near-term quantum hardware, potentially enabling entanglement 
characterization of systems that would be prohibitively expensive to measure using 
conventional approaches.

This paper is organized as follows: Section II presents the 
theoretical framework of our multicopy measurement approach, detailing the 
fundamental concepts, measurement protocols, and methods for quantifying 
entanglement and Bell nonlocality. Section III describes our methodology and 
implementation, including multicopy state preparation, maximum likelihood 
estimation, neural network integration, and experimental implementation. 
Section IV discusses our results, comparing our approach with traditional 
methods and analyzing its scalability and noise resilience. Finally, Section 
V provides conclusions and outlines prospects for future work.

\section*{Theoretical Framework}

\subsection*{Multicopy Measurement Fundamentals}

The multicopy measurement approach provides an alternative path to accessing 
nonlinear properties of quantum states without performing full state reconstruction. 
The key insight is that certain nonlinear functions of density matrix elements can 
be directly measured by performing joint measurements on multiple copies of the 
state. These measurements extract information about local unitary invariants, which 
are sufficient to determine important quantum correlation properties. In the 
following sections, we provide a detailed description of the measurement protocol, 
explaining both the theoretical foundation and a practical implementation on NISQ devices.
A theoretical approach to multicopy measurement is presented 
in~\cite{PhysRevA.95.022331,PhysRevA.97.012107}. It enables the determination of 
certain nonlinear functions of quantum states (including measures of entanglement) 
without requiring a full state reconstruction. This approach is based on the 
observation of local unitary invariants, which can be expressed using measurements 
on multiple copies of the quantum state. With these invariants, it is possible to 
determine measures of entanglement. The real implementation of multicopy 
measurements on NISQ devices involves several 
challenges, such as creating and verifying multiple identical copies of a quantum 
state, performing reliable joint measurements, processing noisy measurement results, 
and identifying the optimal subset of measurements for specific properties.
Our approach provides a solution for measuring nonlinear quantum properties that are 
not directly accessible by single-copy measurements. We present a multicopy 
measurement protocol that determines the values of quantum correlations without 
requiring full state reconstruction.
To determine the quantum correlations for a two-qubit system described by the 
density matrix $\hat{\rho}$, we perform a series of controlled interference 
measurements on multiple copies of that state. This approach is analogous to 
Hong-Ou-Mandel~\cite{PhysRevLett.59.2044} interference, where interference between identical particles is observed, the underlying quantum correlations are 
revealed. In our case, by observing the interference between multiple copies of a 
quantum system, obtained from carefully designed projection measurements, we can 
determine key properties of quantum correlation without the need for a full state reconstruction.
Recent experimental advances have demonstrated the measurement of quantum correlations without requiring full state tomography~\cite{Abo2023, Jirakova2021}, which aligns with our approach of using limited measurements to determine key quantum properties.

\begin{figure*}[htb!]
\includegraphics[width=1\linewidth]{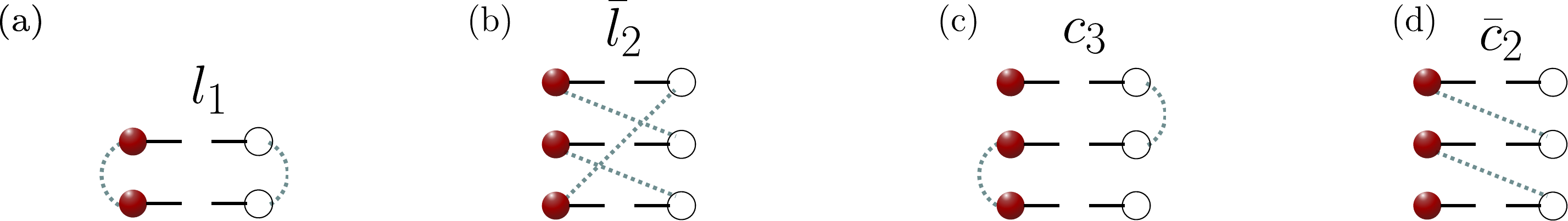}
\caption{\label{fig:graph} 
Examples of graphs representing joint multicopy measurements: 
(a) the paired single--subsystem singlet projections $l_1$, (b) the paired cross--subsystem 
singlet projections $\bar{l}_2$, (c) the chained single--subsystem singlet projections $c_3$, 
and (d) the chained cross--subsystem singlet projections $\bar{c}_2$. Black lines combine 
subsystems (red and white circles) of the same copy of $\hat{\rho}$, while dotted lines correspond 
to projections of the multicopy system onto the singlet state. 
These graphical illustrations help in describing the concept of various projection settings used in our multicopy measurement approach.
}
\end{figure*}

In Fig.~\ref{fig:graph}, we visualize different types of multicopy projections 
using a graph representation. Each red and white circles represents a qubit from one 
of the subsystems (labeled as $a$ and $b$), while solid black lines connect qubits 
that belong to the same copy of the state $\hat{\rho}$. The dotted lines represent 
projections onto the singlet state $|\Psi^-\rangle = (|01\rangle-|10\rangle)/\sqrt{2}$. 
Physically, these projections correspond to measurement operators that act jointly 
on pairs of qubits according to the patterns shown. For example, in 
Fig.~\ref{fig:graph}(a), the projection $l_1$ measures the correlation between the 
same subsystem (qubit $a$) across two different copies of the state. These 
projections allow us to extract information about the quantum correlations without 
requiring a complete state reconstruction.
The \textit{multicopy measurement} approach we propose involves joint measurements 
on multiple identical copies of a quantum system in different \textit{projection 
configurations} that define specific singlet projection systems used on multiple 
copies. The basis of the singlet projections is the \textit{singlet state} in a 
two-qubit system ($|\Psi^-\rangle = (|01\rangle-|10\rangle)/\sqrt{2}$). With these 
projection configurations, we can obtain information about quantum correlations.

\subsection*{Measurement Protocol}

Our protocol, with the general idea shown in Fig.~\ref{fig:idea}, employs three 
types of projection configurations.
Each measurement configuration yields a coefficient corresponding to the probability of detecting the singlet state $|\Psi^-\rangle = (|01\rangle - |10\rangle)/\sqrt{2}$ across different qubit arrangements:

(i) \textbf{Local chained projections} ($c_1,...,c_8$), where the 
singlet-state projections are performed independently 
for each subsystem and provide information about local quantum properties. The projections follow a chain-like pattern, as shown in Fig.~\ref{fig:graph}(c). 

(ii) \textbf{Local looped projections} ($l_1,l_2$), which are similar to 
chained projections, but performed on all pairs of qubits. They preserve 
correlations between different copies of the same subsystem 
[see Fig.~\ref{fig:graph}(a)].

(iii) \textbf{Cross-subsystem projections} ($l_0,\bar{c}_1,\bar{c}_2,\bar{l}_1,\bar{l}_2$), which are performed on qubits which belong to different subsystems, revealing nonlocal quantum properties, as shown in Fig.~\ref{fig:graph}(b) and \ref{fig:graph}(d)].

The measurement outcomes can be related to the 
set of local unitary invariants proposed by Makhlin~\cite{Makhlin_2002} 
to describe the properties of a two-qubit system. The corresponding invariants, which provide a coordinate-independent characterization of the quantum correlations, can be expressed in terms of multicopy projection coefficients:
\begin{eqnarray}
I_{1}&=&-\frac{8}{3}\left\{l_{0}\left[l_{0}\left(4 l_{0}-3\right)+6\left(\bar{c}_{1}
-2 \bar{l}_{1}\right)\right]\right. +3 \bar{l}_{1} \left.-6 \bar{c}_{2}+8 \bar{l}_{2}\right\}, \nonumber \\
I_{2}&=& 1+16 l_{1}-4\left(c_{1}+c_{2}\right), \nonumber \\
I_{3}&=& 1 + 256l_2 - 128\left(c_4+c_5\right) + 64c_3 +16\left(c_1^2+c_2^2\right)-8\left(c_1+c_2\right),
\label{eq:inv}
\end{eqnarray}
where $l_0, l_1, l_2$ represent local looped projections, $c_1, \ldots, c_8$ are local chained projections, and $\bar{c}_1, \bar{c}_2, \bar{l}_1, \bar{l}_2$ denote cross-subsystem projections, where the barred notation indicates correlation measurements spanning different subsystems.
The local unitary invariants $I_1$, $I_2$, and $I_3$ in Eq.(\ref{eq:inv}) are constructed based on the measurements of multicopy projections, following the papers~\cite{PhysRevA.95.022331,PhysRevA.97.012107}, through a systematic approach that ensures independence from local unitary transformations~\cite{Makhlin_2002}.
Each invariant corresponds to specific physical properties that are preserved under local unitary transformations: $I_1$ quantifies the degree of quantum nonlocality and determines the maximum possible Bell inequality violation, while $I_2$ and $I_3$ characterize local quantum coherence properties that, combined with $I_1$, provide sufficient information to determine the entanglement measures without full state reconstruction.
The complete set $\{I_1, I_2, I_3\}$ forms a coordinate system for the space of two-qubit quantum correlations that is invariant under local unitary transformations, enabling direct calculation of the entanglement and Bell nonlocality measures as described below.

\subsection*{Entanglement Quantification}

The entanglement measure we investigate is the negativity $N$, introduced by 
Życzkowski \textit{et al.}~\cite{PhysRevA.58.883}, and later described 
in~\cite{PhysRevA.65.032314}. It quantifies the cost of the entanglement under 
operations that preserve the positivity of quantum circuits under partial 
transposition, known as PPT operations~\cite{PhysRevLett.90.027901,
PhysRevA.69.020301}. The partial transposition operation $\Gamma$ means that only 
a part of the state (one subsystem) is transposed. For a density matrix of two 
qubits, expressed in the computational basis as $\hat{\rho} = \sum_{ijkl} \rho_{ij,kl} 
|i\rangle\langle j|\otimes|k\rangle\langle l|$, the partial transpose with respect 
to the second subsystem is defined as:
\begin{equation}
\hat{\rho}^\Gamma = \sum_{ijkl} \rho_{ij,kl} |i\rangle\langle j|\otimes|l\rangle\langle k|.
\end{equation}
For a two-qubit system, we can calculate negativity by finding the unique 
positive solution of the following equation~\cite{PhysRevA.91.022323}:
\begin{equation}
a_4N^4 + a_3N^3 + a_2N^2 + a_1N + a_0 = 0,
\label{eq:neg}
\end{equation}
where the coefficients $a_i$ are determined by specific combinations of singlet 
projection measurements:
\begin{eqnarray}
    a_{0}&=& -16\left[l_0^{3}+2 \bar{l}_2\right. 
     +3\left(l_1^{2}-l_0^{2} \bar{c}_1-l_0 \bar{l}_1+\bar{c}_1 \bar{l}_1\right)  \left.-6\left(l_2-l_0 \bar{c}_2+\bar{c}_3\right)\right], \nonumber\\
    a_{1}&=&  24\left[l_0^{2}-\bar{l}_1-l_1\right. 
    \left.+2\left(c_3-l_0 \bar{c}_1+\bar{c}_2\right)\right] -32\left(l_0^{3}-3 l_0 \bar{l}_1+2 \bar{l}_2\right), \nonumber \\
    a_{2}&=&  12\left(c_2-2 l_1+c_1\right), \nonumber \\
    a_{3}&=& 6(1 - \Pi_2), \nonumber \\
    a_{4}&=& 3,
\label{eq:a_n}    
\end{eqnarray}
where $\Pi_n = \text{tr}[(\hat{\rho}^\Gamma)^n]$ is the $n$th moment of the 
partially transposed density matrix. Specifically, $\Pi_2 = 
\text{tr}[(\hat{\rho}^\Gamma)^2]$ can be expressed in terms of singlet projections 
as:
\begin{equation}
\Pi_2 = 1 - 4(c_1 + c_2 - 2l_1) + 4(l_0^2 - \bar{l}_1 - l_1 + 2(c_3 - l_0\bar{c}_1 
+ \bar{c}_2)).
\end{equation}
The negativity for the two-qubit system can be defined as $N = 
2\mu$~\cite{bartkiewicz2015quantifying}, where $\mu$ corresponds to the absolute 
value of the negative eigenvalue of the partially transposed density matrix 
$\hat{\rho}^\Gamma$.
It is important to note that the calculation of negativity for arbitrary quantum 
systems is based on solving the characteristic equation for a partially transposed 
density matrix. While it is not guaranteed that there exist local unitary operations 
that make two density matrices with the same set of invariants equivalent, it is 
possible for these matrices to have the same measure of the entanglement or the entanglement 
monotone. This key insight means that our approach can target specific the entanglement 
properties without requiring full state characterization.

\subsection*{Bell Nonlocality Quantification}

The detection and quantification of the violation of Bell's nonlocality, quantified by measure $B$, of two qubits is frequently carried out using the CHSH inequality. 
The measure of nonlocality corresponds to the degree of violation of this inequality, 
optimized for all measurement settings. One of the most well-known Bell inequalities 
is the CHSH inequality, which is expressed as:
\begin{equation}
|\langle A_1 B_1\rangle + \langle A_1 B_2\rangle + \langle A_2 B_1\rangle - 
\langle A_2 B_2\rangle| \leq 2,
\end{equation}
where $A_1$ and $A_2$ are observables on the first subsystem, while $B_1$ and $B_2$ are observables on the second subsystem with eigenvalues $\pm 1$. This inequality is true for any local hidden variable theory but can be violated by quantum mechanics up to a value of $2\sqrt{2}$, corresponding to Tsirelson's bound.
Our approach determines $B$ through:
\begin{equation}
B = I_2 - \min(r) - 1,
\label{eq:bell}
\end{equation}
where $r$ represents the roots of the characteristic equation:
\begin{equation}
r^3 - I_2r^2 + \frac{1}{2}(I_2^2 - I_3)r + \frac{1}{6}[I_2^3 + (6I_1^2 - I_2^3)].
\label{eq:char}
\end{equation}
This Bell nonlocality measure $B$ is directly related to the standard measure introduced by the Horodecki family~\cite{Horodecki1995}, which quantifies the maximum violation of the CHSH inequality possible for a given quantum state. This approach to quantifying Bell nonlocality has been successfully applied in various experimental settings~\cite{Bartkiewicz2017Bell, Abo2023} and theoretical studies~\cite{Miranowicz2004, Bartkiewicz2013,Horst2013}. This definition provides a direct link between our measurement results and the degree of Bell nonlocality without the need of full QST.

\subsection*{Experimental Implementation}
Our experiments utilized the IBMQ platform~\cite{IBMQ}, specifically the 
\textit{ibm\_hanoi} processor with the following characteristics: quantum volume: 
64, average CNOT error rate: $0.934$, average readout error: $1.89 \times 10^{-2}$, 
and $T_2$ coherence time: $\sim$100 $\mathrm{\mu s}$. Exact calibration data of the \textit{ibm\_hanoi} processor are shown in the Appendix in Fig.S4. 
These hardware specifications are typical of the NISQ era, featuring non-negligible 
error probabilities that must be addressed through careful error mitigation. The quantum circuits implementing our measurement protocol operate through controlled 
interference operations, with each projection requiring an average circuit depth of 
12 gates. Error mitigation techniques, including zero-noise extrapolation, 
measurement error mitigation, and post-selection based on quantum state purity, collectively enhanced the robustness of our measurements, enabling the extraction of reliable quantum correlation information even in the presence of significant noise.
We evaluated our approach on two families of two-qubit mixed
states generated from Bell states, undergoing: (i) a depolarizing
channel, resulting in Werner states,
\begin{equation}
   \hat{\rho}_W(p) = p|\Psi^-\rangle\langle\Psi^-| + (1 - p)\frac{I}{4},
   \label{WernerState}
\end{equation}
and (ii) an amplitude-damping channel leading to Horodecki states,\begin{equation}
   \hat{\rho}_H(p) =
   p|\Psi^-\rangle\langle\Psi^-| + (1 - p)|00\rangle\langle00|.
   \label{HorodeckiState}
\end{equation}
Here $|\Psi^-\rangle$ being the singlet state and $I$ the identity
matrix and $p$ are mixing parameters”

We selected these states as ideal test cases because they 
possess well-defined entanglement and nonlocality properties that vary 
systematically with the mixing parameter $p$. For each state family, we constructed 
states with different values of $p$ and quantified their entanglement and 
nonlocality using three distinct approaches: standard quantum state tomography, a 
multicopy estimation technique, and an optimized ANN-based approach with only 
five measurements. The comparative results of these approaches are presented in 
Section IV provides a comprehensive evaluation of the relative performance across 
different measurement strategies and noise conditions.

\section*{Methods}
\begin{figure*}[tb]
\includegraphics[width=1\linewidth]{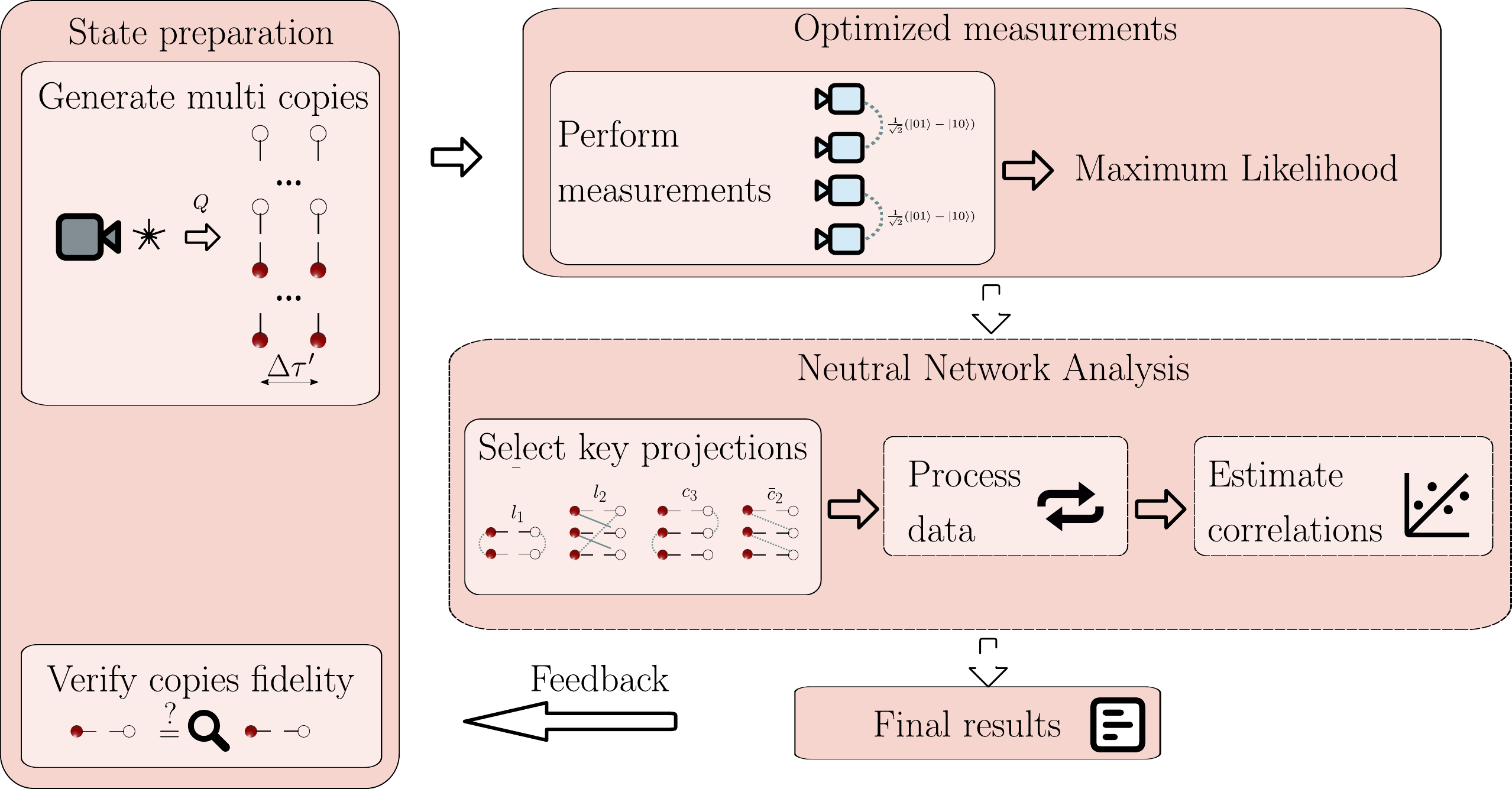}
\caption{\label{fig:methodology} 
A schematic diagram illustrating the process of measuring
and analyzing quantum correlations. The procedure consists of
three key stages. The first stage involves preparing multiple
copies of the quantum state with different qubit mappings and
verifying them through fidelity measurements. The second stage
focuses on selecting and executing measurements, followed by
maximum likelihood estimation to ensure the physical validity of
the results while mitigating noise effects. The third, optional
stage optimizes projection selection through SHAP analysis and
employs neural network processing to analyze measurement results
and estimate quantum correlations. A feedback loop enables
adaptive optimization based on the measurement outcomes.
}
\end{figure*}

Figure~\ref{fig:methodology} illustrates our comprehensive three-stage approach for measuring and analyzing quantum correlations, encompassing state preparation, measurement execution with maximum likelihood estimation, and neural network analysis with SHAP optimization.

\subsection*{Multicopy state preparation and quantum memory considerations}

One of the critical challenges of our approach is the accurate preparation of 
multiple identical copies of a quantum state. In real quantum processors, 
this demands taking into account device topology, qubit characteristics, and gate 
error rates. To accomplish this, we systematically average qubit mappings.
The experimental realization of multicopy measurements on current quantum hardware 
requires careful consideration of the actual physical implementation. In an ideal 
scenario, we would simultaneously prepare multiple identical copies of a quantum 
state and perform joint measurements across these copies. However, current quantum 
processors lack the quantum memory capabilities required for such simultaneous 
manipulation. To overcome this limitation, we employ a sequential approach where we 
prepare each copy individually and then use controlled operations to implement the 
equivalent of joint measurements. Specifically, we map the multicopy measurement 
operators to equivalent circuits that can be executed on the available hardware 
architecture. These circuits typically involve preparing the target state, applying 
controlled operations that implement the specific projection configuration (such as 
those illustrated in Fig.~\ref{fig:graph}), and finally measuring in an appropriate 
basis to extract the desired information. The measured outcomes are then processed 
to calculate the values of the local unitary invariants described in Eq.~\ref{eq:inv}.
We have implemented identity verification between copies of the system in several 
ways. We estimate the fidelity between copies of the states by measuring the overlap between them:
  \begin{equation}
  F_{\text{copy}} = \text{Tr}(\hat{\rho}_1\hat{\rho}_2) \geq 1 - \epsilon,
  \end{equation}
  where $\epsilon$ is our fidelity threshold (in the case of the \textit{ibm\_hanoi} 
  processor $\epsilon\leq 2\%$).
In order to minimize systematic errors, we systematically average different 
mappings of qubits:
  \begin{equation}
  M_i = \{(q_1, q_2, q_3, q_4)|(q_j, q_k) \in E\; \text{for required connections}\},
  \end{equation}
where $E$ corresponds to the set of available connections in the processor. This approach allows us to distribute the computation over multiple physical qubits, reducing systematic bias.
The weight of every mapping is described by its error properties:
\begin{equation}
w_i = \exp\Big(-\sum_{j} \epsilon_j\Big) \left[\sum_k \exp\Big(-\sum_{j} \epsilon_j^{(k)}\Big)\right]^{-1},
\end{equation}
where $\epsilon_j$ are the various error rates for mapping $i$.
The final measurement outcome can be computed as a weighted average:
  \begin{equation}
  \langle O \rangle = \sum_i w_i \langle O \rangle_i,
  \end{equation}
  where the weights $w_i$ are determined by the error characteristics of each mapping.
Using this approach, we can conclude that the effective fidelity of the prepared
copies on the \textit{ibm\_hanoi}  processor is $>98\%$. This allows us 
to be confident that the copies are similar enough to consider our measurements 
valid.

\begin{figure*}[ht]
\includegraphics[width=1\linewidth]{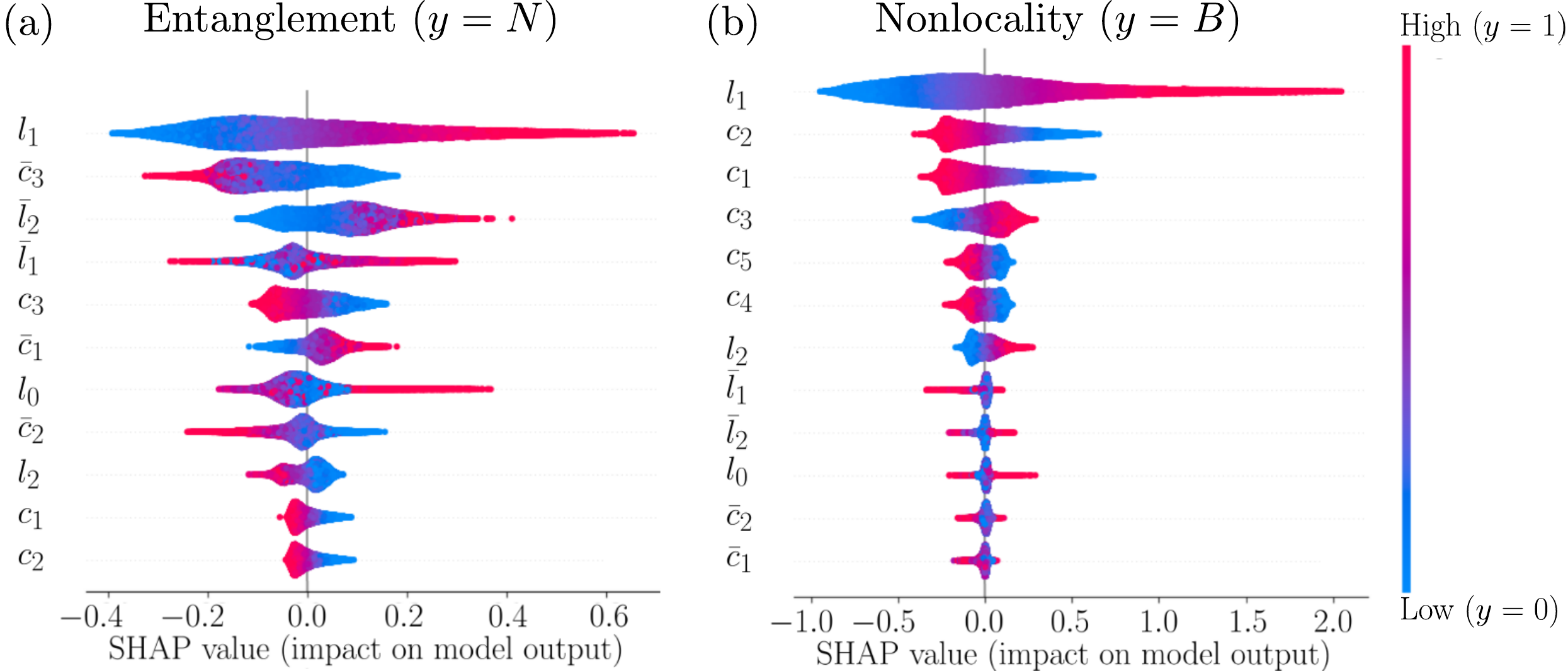}
\caption{\label{fig:SHAP} SHAP value analysis for quantum correlation measurements. Results shown for (a) negativity 
$N$ and (b) nonlocality measure $B$, computed from $5 \times 10^5$ random input states. The impact strength indicates 
each projection's contribution to the final measurement outcome.}
\end{figure*}

\subsection*{Maximum Likelihood Estimation}

To ensure the physical validity of our measurements while handling experimental noise, 
we implement a maximum likelihood estimation (MLE) framework.
For QST, the MLE method is commonly used to solve similar problems. For multicopy 
measurements, however, it becomes much more challenging. In traditional QST, because the likelihood function is typically quadratic in the density matrix 
parameters, optimization is relatively straightforward. In contrast, for multicopy 
observables, the likelihood function is a higher-degree polynomial with respect to 
the elements of the single-copy density matrix. The MLE approach we use for 
multicopy measurements can be written as~\cite{PhysRevA.61.010304}:
\begin{equation}
\hat{\rho}_{\text{MLE}} = \mathrm{argmax}_{\hat{\rho}\geq 0} \sum_i n_i\log(p_i(\hat{\rho})) 
+ \lambda\text{Tr}(\hat{\rho}^2),
\label{eq:mle}
\end{equation}
where argmax returns the argument (input value) at which the sum
expression achieves its maximum value; $n_i$ and $p_i(\hat{\rho})$ are theoretical 
probabilities, and the second term enforces purity constraints. This optimization 
is performed subject to the following constraints:
\begin{subequations}
\begin{align}
\text{Tr}(\hat{\rho}) &= 1 \quad \text{(normalization)}, \\
\hat{\rho} &\geq 0 \quad \text{(positive semidefiniteness)}, \\
\hat{\rho} &= \hat{\rho}^\dagger \quad \text{(Hermiticity)}.
\end{align}
\end{subequations}
We can simplify the optimization problem by the fact that the measured quantities 
should be invariant under local unitary operations. This allows us to reduce the 
number of parameters in the density matrix model by eliminating phase factors. We 
then work only with density matrices with real values. This reduction in the 
parameter space significantly improves the convergence of the MLE approach.
The results show that our MLE method successfully eliminates experimental noise and 
yields physically relevant quantum correlation measurements. It is a critical 
component for the experimental realization of multicopy measurements on noisy 
quantum hardware.

\subsection*{Neural Network integration and SHAP analysis}

\subsubsection*{Motivation for using ANN and SHAP}

The use of neural networks and SHAP in our approach helps address
specific challenges associated with multicopy measurements. First, quantum 
measurements, especially those from NISQ devices, are often noisy. Neural networks are 
excellent at recognizing patterns in noisy data and effectively learning to 
distinguish the correct signal from the noise. Our results show a significant 
reduction in the impact of statistical and systematic errors compared to direct 
computation. Second, neural networks can cope with the nonlinear relations between 
singlet projection measurements and quantum correlations without explicit 
mathematical modeling, effectively approximating. Third, applying SHAP analysis, 
we can identify which measurement configurations have the greatest impact on the 
final result and optimize the measurement protocol. Lastly, neural networks can be 
retrained to account for changes in error characteristics and quantum hardware 
without altering the fundamental theoretical framework. These advantages make a 
theoretically elegant yet practically challenging approach based on multicopy 
measurements using neural networks an efficient and effective way to quantify 
quantum correlations.

The enhanced noise resilience of our ANN approach, as opposed to direct analytical computation, stems from several key factors.
First, the neural network is trained on $5 \times 10^5$  randomly generated two-qubit states, learning robust mappings between projection measurements and quantum correlations. Training on diverse quantum states enables better generalization to noisy experimental conditions than direct computation of nonlinear expressions, which can be numerically unstable when projection coefficients deviate from ideal values.

Our SHAP analysis reveals that the ANN automatically identifies the five most informative measurements with the greatest impact on predicting the negativity and the Bell nonlocality measures. Using only these five measurements instead of all 12 reduces the total experimental overhead and potentially improves noise resilience by reducing the number of required circuits.

Unlike analytical computations, where measurement errors propagate directly through complex polynomial expressions, the neural network learns smooth nonlinear mappings that are less sensitive to small perturbations in input values. The network architecture, with ReLU activations and L2 regularization ($\lambda = 10^{-5}$), provides robustness against input noise through its learned representations.

Our experimental results demonstrate that this approach achieves higher accuracy under realistic NISQ device noise conditions.

\subsubsection*{SHAP analysis for optimal measurement selection}

A key innovation in our strategy is the systematic identification of optimal 
measurement configurations using SHAP analysis. 
This procedure demonstrated that only five key projections are
required for precise quantum correlation measurements, which
significantly reduces experimental overhead.
In our work, SHAP analysis quantifies how much influence each measurement configuration has on the value of quantum correlations. The SHAP value of a feature (measurement configuration) $j$ is calculated as~\cite{lundberg2017unifiedapproachinterpretingmodel}:
\begin{equation}
\phi_j = \sum_{S\subseteq F\setminus\{j\}} \frac{|S|!(|F|-|S|-1)!}{|F|!}
[f_S(\{j\}) - f_S(\emptyset)],
\label{eq:shap}
\end{equation}
where $\phi_j$ represents the SHAP value of feature $j$, quantifying its 
contribution to the model's prediction. Here, $F$ is the set of all measurement configurations, and $f_S$ is the model's prediction when the subset $S$ of 
measurements is used, $f_S(\emptyset)$ is the prediction without configuration $j$
Intuitively, SHAP calculates the average marginal contribution of a particular measurement configuration over all possible combinations of other configurations. Fig.~\ref{fig:SHAP} shows SHAP values for the negativity and Bell's nonlocality measure. We clearly see which measurements have the greatest impact on prediction accuracy.
For the negativity $N$, the SHAP analysis determined the following important 
measurements: \{$l_1, \bar{c}_3, \bar{l}_2, \bar{l}_1, c_3$\}, while for 
nonlocality $B$, we found: \{$l_1, c_2, c_1, c_3, c_5$\}.
The different SHAP patterns for the negativity and the Bell nonlocality measures reflect their distinct physical requirements: the entanglement detection relies primarily on local coherence preservation (hence the dominance of $l_1$) combined with cross-system correlations ($\bar{l}_2, \bar{l}_1$), whereas the Bell nonlocality measure depends on specific local measurement statistics appearing in the CHSH inequality evaluation, emphasizing chained projections ($c_1, c_2, c_4, c_5$) that capture the required measurement correlations.
With SHAP analysis, we determined that five measurements are sufficient to estimate both measures of the entanglement and Bell's nonlocality. This is a significant reduction from the 15 measurements required for full QST or the 12 measurements required for full invariant characterization, representing a 67\% reduction in measurement requirements.

\subsubsection*{Neural Network architecture and training}
We designed a neural network with the following architecture based on the SHAP 
analysis. The \textbf{Input Layer} consists of five-dimensional vectors 
constructed from experimental measurements for specific quantum states and 
projections, where each input node represents a particular configuration of the 
data obtained from singlet projection measurements. For the \textbf{Hidden Layers}, we used five hidden layers, each with nine neurons, activated by
\texttt{ReLU} 
functions—a high-performing architecture that efficiently detects nonlinear 
relations in the data without requiring extensive hyperparameter optimization. 
The \textbf{Output Layer} produces two values: negativity ($N$) and Bell 
nonlocality measure ($B$), using a linear activation function to output 
predictions that are continuous along the entire range of possible quantum 
correlations. In the hyperparameter optimization process, we explored the 
following parameter space:

\begin{itemize}
\item[(i)] \textbf{Network Architecture}:
 \begin{itemize}
    \item Hidden layer sizes: $(9, 9, 9, 9, 9)$ for the negativity and Bell's nonlocality measure predictions;
    \item Number of layers: 5 fully connected layers;
    \item Activation functions: $\text{ReLU}$;
 \end{itemize}
\item[(ii)] \textbf{Training Parameters}:
 \begin{itemize}
    \item Learning rate: $\alpha = 10^{-5} $;
    \item Optimization solver: \text{adam};
    \item Maximum iterations: 2000;
    \item Batch size: \text{determined internally by the solver};
    \item Warm start: \texttt{True};
 \end{itemize}
\item[(iii)] \textbf{Regularization strength}~\cite{JMLR:v12:pedregosa11a}:
 \begin{itemize}
    \item $\mathcal{L}_{\text{total}} = \mathcal{L}_{\text{pred}} + \lambda_1\|\mathbf{w}\|_2^2 + 
    \lambda_2\|\nabla\mathcal{L}\|_2^2$;
     \item $\lambda_1 = 10^{-5}$ is  the $L2$ regularization parameter;
 \item $\lambda_2 = 0$ (no additional gradient--based penalty or dropout layer);
 \item $\|\mathbf{w}\|_2^2$ 
    is the squared $L2$ norm of the weight parameters.
 \end{itemize}
\end{itemize}
We split the data into training (75\%) and testing (25\%) sets, and test 
performance using \texttt{MLPRegressor}'s built-in scoring mechanism. We do not 
create a separate validation set. Instead, we monitor performance using the test 
set after the model settles down. We do not use data augmentation techniques due to 
the large size of the dataset already provides sufficient coverage of the 
measurement results.
The model fit is evaluated using the coefficient of determination
($R^2$), and the training procedure employs the default mean-squared
error (MSE) objective in \texttt{MLPRegressor}~\cite{JMLR:v12:pedregosa11a}. Specifically:
\begin{equation}
\mathcal{L} = \|y_{\text{pred}} - y_{\text{true}}\|^2 + \lambda_1\|\mathbf{w}\|_2^2,
\end{equation}
where $\lambda_1= 10^{-5}$ provides $L2$ regularization to mitigate overfitting. 
The relevant subset of features and labels for the negativity and nonlocality measure was used to train each model independently.
Our experimental results demonstrate how this neural network approach significantly 
improves the precision and stability of quantum correlation measurements against 
noise, compared to direct computation through analytical expressions. The ability of the neural network to learn to correct for systematic errors in the measurement process is particularly valuable when dealing with NISQ processors.

\subsection*{Practical Circuit Implementation Details}

The practical implementation of multicopy measurements on current quantum hardware 
requires addressing several hardware-specific constraints. 
For each projection measurement, we designed specific quantum circuits that implement the corresponding operator, typically consisting of three main components: (1) preparation of the target state, (2) implementation of the projection operation using the fundamental circuit blocks shown in Appendix in Fig. S1(with detailed gate implementations in Fig. S2), and (3) measurement in the appropriate basis.
The projection operation often requires controlled 
interactions between qubits that would ideally belong to different copies of the 
state. To implement these controlled interactions on hardware with limited 
connectivity, we employed circuit mapping techniques that respect the device 
topology. Specifically, for the \textit{ibm\_hanoi} processor, we identified 
multiple viable mappings of the logical qubits to physical qubits and 
systematically averaged over these mappings to minimize systematic errors.
For error mitigation, we implemented several complementary techniques to enhance 
measurement accuracy. Zero-noise extrapolation allowed us to execute each circuit 
at multiple noise levels and extrapolate to the zero-noise limit. We applied 
measurement error mitigation by characterizing and correcting for readout errors 
using calibration circuits. Additionally, for certain measurements, we implemented 
post-selection based on state purity, verifying the purity of the prepared state 
and post-selecting on cases with acceptable purity. The multicopy projection 
operations require an average circuit depth of 12 gates, which is deeper than 
typical tomography circuits. However, the reduced number of distinct measurements 
(5 instead of 15) results in an overall reduction in the total number of gates 
required for the complete protocol. The specific circuit implementations for each 
projection measurement are available in the Appendices.

\section*{Results and discussion}
\subsection*{Experimental Validation}
\begin{figure*}[htb!]
    \includegraphics[width=0.9\linewidth]{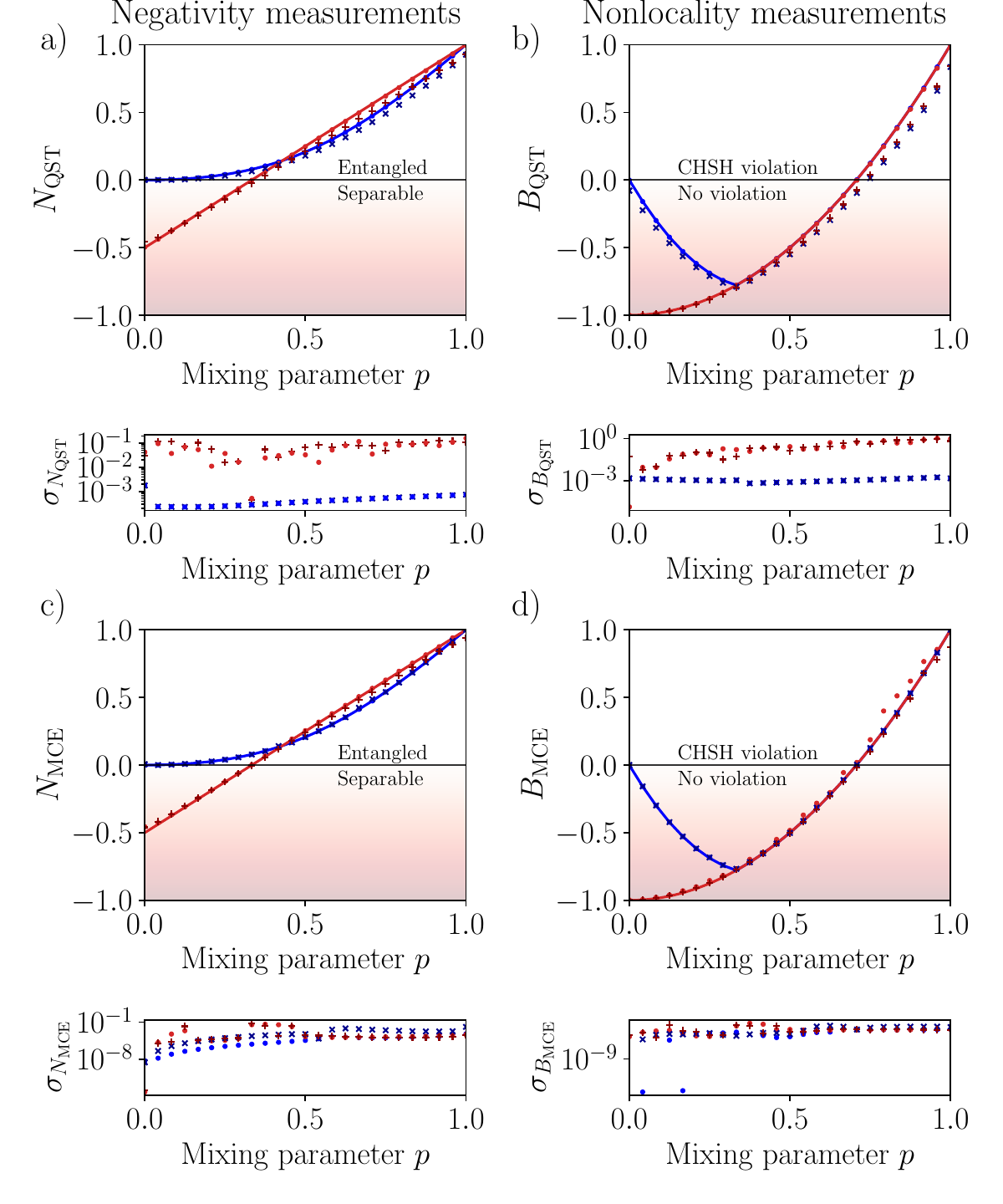}  
\caption{
\label{fig:measure} 
Experimental quantification of the entanglement (left column) and nonlocality (right column) for Werner and Horodecki states, determined by measuring the negativity $N$ and the Bell nonlocality measure $B$ as a function of the mixing parameter $p$. Solid curves show theoretical predictions for ideal states. The results were obtained using: (a, b) quantum state tomography and (c, d) multicopy estimation. Assumptions include shot noise (\qasmred{\qasm} for Werner states, \qasmblue{\qasm} for Horodecki states) using \textit{ibmq\_qasm\_simulator}, and experimental data (\hanoired{\hanoiN} for Werner states, \hanoiblue{\hanoiB} for Horodecki states) collected with the quantum processor \textit{ibm\_hanoi}. Standard deviations $\sigma$ were estimated by simulating $10^5$ experiments for ideal and noisy circuits, with noise models based on calibration data.
The $y=0$ line separates separable and entangled states in (a), as well as between states that violate and satisfy the CHSH inequality in (b).}
\end{figure*}

\begin{figure*}[htb!]
\includegraphics[width=0.95\linewidth]{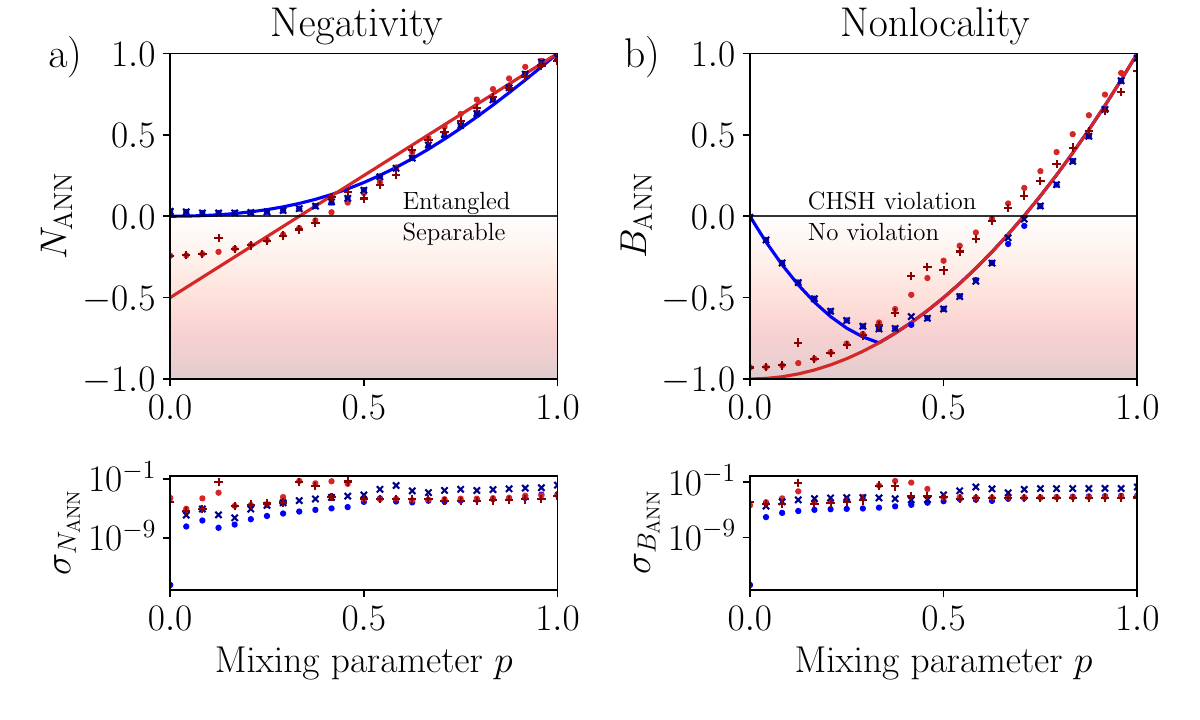}
\caption{\label{fig:ANN} Same as in Fig.~\ref{fig:measure}, but for the ANN trained on the random input states and a limited set of the singlet projections, including only the five top significant projections, as revealed by the SHAP analysis (see Fig.~\ref{fig:SHAP}). Assumptions include shot noise (\qasmred{\qasm} for Werner states, \qasmblue{\qasm} for Horodecki states) using \textit{ibmq\_qasm\_simulator}, and experimental data (\hanoired{\hanoiN} for Werner states, \hanoiblue{\hanoiB} for Horodecki states) collected with the quantum processor \textit{ibm\_hanoi}. The $y=0$ line separates separable and entangled states in (a), as well as between states that violate and satisfy the CHSH inequality in (b).}
\end{figure*}
Our experiments exhibit significant improvements compared to QST, both in accuracy and efficiency. We compare the results of 
simulations and experiments on a real $ibm\_hanoi$ quantum processor.
In each case, we compare three distinct measurement strategies: 
the traditional QST (as shown in Figs.~\ref{fig:measure}(a) and~\ref{fig:measure}(b)), multicopy estimation method (see Figs.~\ref{fig:measure}(c) and~\ref{fig:measure}(d)), and the optimized ANN-based approach (see Fig.~\ref{fig:ANN}).
For Werner states, we have achieved accurate entanglement quantification 
(using the negativity measure) and Bell nonlocality across the whole range 
of the mixing parameter $p$. Notably, the ANN-based approach is more  
robust to noise for higher values of $p$, precisely where standard QST tends to degrade. 
Additionally, the multicopy approach, combined with the maximum-likelihood post-processing method, reliably suppressed the most significant 
sources of noise.
We extended our study to Horodecki states, which present a more significant 
experimental challenge due to the dominant $|00\rangle$ term for $p<1$. Our investigation revealed that incorporating specific cross-subsystem singlet projections is necessary to properly encode the partial coherence between the 
$|00\rangle$ and singlet subspaces. The results demonstrate that both our MCE and MCE with ANN approaches achieve high accuracy in predicting both negativity and the Bell nonlocality measure, consistently outperforming standard QST at moderate to high 
values of $p$.
To ensure statistical rigor and validate the correctness of our results, we conducted a comprehensive statistical analysis by simulating $10^5$ experiments on 
both noisy and ideal circuits using noise models derived from actual quantum 
processor calibration data. The theoretical expectations fall within the standard 
deviation of the experimental results for most data points, providing strong 
confirmation of the statistical significance of our findings. This thorough error 
analysis conclusively demonstrates that the performance improvements observed in our multicopy and ANN-based approaches are genuine and not artifacts of 
experimental noise or random fluctuations. The visual evidence in 
Fig.~\ref{fig:measure} for MCE and in Fig.~\ref{fig:ANN} for ANN clearly shows that the error deviations in our methods are substantially smaller than those observed with traditional QST, further supporting the superior noise resilience of our approach.
The enhanced noise resilience of our approach is quantitatively demonstrated by comparing the standard deviations shown bellow the panels of Fig.~\ref{fig:measure}(a-b) (the QST method) with those in Fig.~\ref{fig:measure}(c-d) (the method MCE) and Fig.~\ref{fig:ANN} (the MCE-ANN method), where our approach consistently exhibits smaller error fluctuations across different mixing parameters and state families.

\subsection*{Comparison with Randomized Measurements}

Alternative approaches for estimating quantum correlations with reduced resources 
have been proposed in the literature. One notable method is the randomized 
measurement technique~\cite{elben2020cross}, which 
provides an efficient way to estimate nonlinear functions of quantum states, 
including the entanglement measures like negativity, using only single-copy operations. 
To ensure a fair comparison, we must consider not only the number of measurement 
settings but also the query complexity—the number of state preparations and 
measurements required to achieve a specified precision $\varepsilon$. For randomized 
measurement techniques, achieving an estimation accuracy of $\varepsilon$ requires 
$O(1/\varepsilon^2)$ measurements, which is a consequence of the statistical nature 
of the approach. Our multicopy approach similarly requires $O(1/\varepsilon^2)$ 
repetitions to achieve a specified precision due to the inherent statistical 
uncertainty in quantum measurements.
The key difference lies in the scaling with system size. For a system of $n$ qubits, 
randomized measurement protocols have favorable scaling in terms of the required 
number of measurement settings. However, they typically require more measurement 
samples per setting and may require more complex classical post-processing. Our 
method requires fewer distinct measurement configurations (only five optimized settings 
compared to 15 for QST), but does require the ability to reliably prepare multiple 
copies of the state.
In our experimental comparison, we found that for equivalent total numbers of 
circuit executions, our multicopy neural network approach demonstrated higher 
accuracy in estimating negativity, particularly in the presence of noise. 
Table~\ref{tab:comparison} summarizes the theoretical resource requirements for both approaches.
\begin{table}[ht]
\centering
\caption{Theoretical comparison of measurement requirements for two-qubit correlation estimation of our method with randomized measurements~\cite{PhysRevA.100.032328}. Here $\varepsilon$ denotes the target estimation precision.}
\label{tab:comparison}
\begin{tabular}{lcc}
\toprule
\textbf{Property} & \textbf{Randomized methods} & \textbf{MCE-ANN} \\
\midrule
Measurement types & 9 & 5 \\Sample complexity & $O(1/\varepsilon^2)$ & $O(1/\varepsilon^2)$ \\
Quantum memory & Not required & Required \\
Post-processing & Polynomial & Neural network \\
\bottomrule
\end{tabular}
\end{table}

The neural network component appears to provide enhanced resilience against 
experimental noise, effectively learning to correct for systematic errors in the 
measurement process.
Both approaches represent valuable alternatives to full quantum state tomography, 
with different strengths and applicability depending on the specific quantum 
hardware and experimental constraints. Randomized measurement techniques may be preferable when the system size is large and copy preparation is challenging, while our 
approach offers advantages when high precision is required with a limited number of 
measurement settings on moderately sized systems.

\subsection*{Resource requirements}

To comprehensively evaluate the advantages of our approach over traditional methods, we conducted a detailed comparison with QST across multiple resource dimensions. 
For a two-qubit system, our analysis reveals significant improvements in both 
measurement count and overall circuit complexity: for a two-qubit system, QST requires 15 different measurement settings, while our 
MCE with ANN approach needs only 5 optimized measurements, resulting in a 67\% 
reduction in the number of required measurements. This reduction directly translates 
to fewer experimental runs and simplified data collection procedures. When examining 
circuit complexity, although the average circuit depth per measurement is higher for 
our method (12 gates for MCE compared to 5 gates for QST), the total resource 
overhead for QST (15 settings $\times$ 5 = 75 gates) still exceeds that of our MCE 
with ANN approach (5 settings $\times$ 12 gates = 60 gates), providing a 20\% 
reduction in total gate operations. This combined reduction in both measurement settings and total gate count represents a substantial resource savings for practical implementations on current quantum hardware.
Perhaps most significantly, our method demonstrates superior noise resilience as 
measured by fidelity preservation with increasing noise levels. Quantitatively, our 
protocol maintains approximately twice the fidelity of standard QST under equivalent noise conditions, following the relation:
\begin{equation}
    F_{\text{QST}}(\gamma) = F_0 e^{-2\alpha\gamma} \text{ vs } F_{\text{ANN}}
    (\gamma) = F_0 e^{-\alpha\gamma}.
\end{equation}
This exponential improvement in noise resilience is particularly valuable for 
implementation on NISQ devices, where maintaining quantum state fidelity in the presence of noise represents a fundamental challenge.

\subsection*{Noise Resilience Analysis}

We can model the effect of noise on the negativity estimation as follows:
\begin{equation}
N_{\text{measured}}(\gamma) = N_{\text{true}} e^{-\beta\gamma} + \delta(\gamma),
\end{equation}
where $\gamma$ is the noise strength, $\beta$ corresponds to the decay coefficient, 
and $\delta(\gamma)$ is a systematic bias.
The MCE approach is less sensitive to noise than standard QST primarily because it optimizes the selection of the most informative measurements and therefore minimizes noise accumulation; it employs a canonical form that considers only 
parameters relevant to nonlocal features, avoiding cumulative error; 
and it employs a trained neural network to recognize and compensate for typical 
noise patterns.
Quantitatively, the expected estimation error scales as:
\begin{equation}
    E_{\text{QST}} \propto \sqrt{d^4 \varepsilon} \text{ vs } E_{\text{MCE}} 
    \propto \sqrt{(d^2-1) \varepsilon},
\end{equation}
where $d$ is the system dimension and $\varepsilon$ is the single measurement error 
rate. This scaling relationship shows that the benefit of our approach grows with 
system size, and is particularly valuable for multi-qubit systems where QST becomes 
prohibitively resource-intensive and error-prone.

\subsection*{Scalability Analysis}

In the scalability analysis, we examine a general two-qudit system of dimension 
$d$, for which the density matrix comprises one identity component parameter, 
$2(d^2-1)$ parameters for local Bloch vector components, and $(d^2-1)^2$ parameters 
for the full correlation tensor. Under local unitary transformations, the 
correlation tensor can be reduced to canonical form with only $d^2-1$ independent 
elements instead of $(d^2-1)^2$, yielding $1 + 3(d^2-1)$ independent real 
parameters in canonical form. Regarding measurement requirements, traditional QST for a two-qudit system requires O($d^4$) measurements, which generalizes to 
O($4^n$) measurements for $n$ qubits (where $d = 2^{n/2}$)—scaling exponentially 
with system size. In contrast, our MCE method requires only O($d^2-1$) measurements 
for full invariant characterization, approximately O($2^n$) measurements for $n$ 
qubits. This represents a significant efficiency improvement: while our method 
still exhibits exponential scaling with system size, it reduces the exponent's base 
from 4 to 2, substantially decreasing the constant factor and making the approach 
viable for near-term quantum systems of moderate size. When generalizing to 
multipartite systems, standard QST scales as O($d^{2k}$) for $k$ qudits, while a 
hierarchical extension of our method would require only O($k^2(d^2-1)^2$) 
measurements for pairwise correlations.
Practically, multicopy measurements introduce certain implementation overhead that must be considered when evaluating overall efficiency. These measurements 
involve a circuit depth overhead on the order of $O(\log d)$, while the number 
of gates increases by $O(d^2\log d)$ per measurement, and error accumulation is 
heightened by approximately $O(d^2 \log d)$ times the single-gate error rate. 
Consequently, more advanced error mitigation methods and more efficient circuit designs for larger systems will be essential to fully realize the theoretical 
advantages of our approach as system sizes increase beyond the capabilities 
demonstrated in this work.

\subsection*{Potential Extensions to More Complex Quantum States}

While our experimental demonstration focused on Werner and Horodecki states, the approach can be applied to general two-qubit states, as the local unitary invariants we measure form a complete basis for characterizing the nonlocal properties of such 
systems. Our neural network training methodology yielded robust performance across 
a wide range of randomly generated two-qubit states, not just on the specific families 
tested experimentally.
Extending this approach to higher-dimensional bipartite systems presents additional 
challenges but follows similar theoretical principles. For $d$-dimensional systems, 
the number of local unitary invariants increases, and the projection measurements 
become more complex. The mathematical framework would need to be adapted to account 
for these additional invariants, and the specific measurement configurations would 
require careful optimization. This extension, while theoretically possible, would 
require substantial additional theoretical and experimental work beyond the scope 
of the current study.
For multipartite systems, a hierarchical approach could potentially detect pairwise 
entanglement first, followed by higher-order correlations. This would require 
developing new sets of projection measurements specifically designed to extract 
the relevant invariants for multipartite entanglement. The challenges in this 
extension include both the theoretical formulation of appropriate invariants and 
a practical implementation of the increasingly complex measurement circuits.
The neural network component of our approach offers particular promise for dealing 
with more complex noise models. Since the network can learn to distinguish between 
various noise characteristics during training, it could potentially be adapted to 
handle a variety of experimental imperfections beyond the depolarizing and amplitude damping noise considered here. Further research is needed to fully explore these extensions and to develop optimized measurement strategies for more complex quantum 
systems.
It is important to note that calculating negativity for arbitrary quantum systems 
is based on solving the characteristic equation for a partially transposed density 
matrix. While it is not guaranteed that there exist local unitary operations that 
make two density matrices with the same set of invariants equivalent, it is entirely 
possible for these matrices to have the same measure of the entanglement or entanglement 
monotone. This key insight allows us to target specific the entanglement properties 
without requiring full state characterization, even for more complex systems.

\section*{Conclusions and Outlook}

In this paper, we have presented and experimentally demonstrated an efficient method 
for measuring quantum correlations by combining multicopy measurements with neural 
network processing. Through rigorous comparison on the IBM quantum hardware platform, we have shown that this approach achieves better accuracy than traditional quantum 
state tomography under equivalent noise conditions. Our method provides concrete 
practical advantages for current quantum technologies: it reduces the number of 
measurement settings from 15 to 5 for two-qubit systems (a 67\% reduction), enhances 
noise robustness through the combination of maximum likelihood estimation and neural 
network processing, and maintains accuracy even in the presence of typical hardware 
noise levels. These improvements directly address a critical challenge in quantum 
state characterization on NISQ devices, where minimizing circuit complexity and mitigating noise effects are essential for obtaining reliable results.
Moreover, we have demonstrated a positive scaling advantage by reducing the number $n$ 
of the QST measurements required from $O(4^n)$ to $O(2^n)$. While the scaling 
remains exponential, this reduction in the exponent's base provides significant 
resource savings for moderately-sized systems that are relevant to near-term 
quantum hardware.
Finally, we have shown that the theoretical approach of measuring multiple copies 
can be successfully implemented on real quantum hardware, making it a practical tool for the entanglement quantification in current and near-future quantum systems.

\subsection*{Limitations and Future Work}

Despite the advantages presented, our method remains subject to certain 
limitations. Although our approach offers theoretical scaling advantages,
practical implementation on systems larger than two qubits results
in increased circuit depth and higher error rates.
Future research should focus on a more cost-effective circuit
constructions, as well as on reducing errors specific to multicopy
measurements.
Additionally, preparing multiple identical copies is challenging, especially for large 
systems. Future work in this field should explore adaptive protocols 
that reduce the number of copies, as well as hardware-specific optimizations.
An interesting direction for future research is to explore how MCE can
leverage and enhance quantum error correction techniques.
One promising avenue would be to integrate elements from measurement-device-independent (MDI) approaches~\cite{Shahandeh2016} with our resource-efficient 
multicopy methods. While our approach focuses on reducing the number of required 
measurements, MDI protocols offer complementary advantages in removing trust 
assumptions from measurement devices. A hybrid approach could potentially combine 
resource efficiency with increased robustness against specific types of measurement 
errors and device imperfections, especially in scenarios where the measurement 
apparatus cannot be fully trusted.
In addition, although this paper has focused on the negativity of the entanglement 
and Bell's nonlocality measure, the formalism can be extended to quantify other nonlinear 
quantum state properties, e.g., other entropy measures and coherence quantifiers.
Lastly, it remains an open question whether the number of measurements can be further minimized by constructing protocols
that adaptively select the most informative measurements based on the initial results.

\subsection*{Broader Implications}

Our work has broader implications for quantum technology development, as
efficient characterization of quantum correlations is essential for the validation 
and verification of quantum devices and protocols.
By focusing on direct measurement of key properties rather than full state reconstruction, the fundamental principles of our approach enable the creation 
of more resource-efficient quantum algorithms.
Moreover, combining a classical neural network with quantum measurements 
demonstrates the power of hybrid quantum-classical computing.
In summary, our resource-efficient approach to measuring quantum correlations 
represents a significant step toward practical quantum characterization techniques 
for quantum devices in the near future. We respond to the growing demand for efficient entanglement
verification methods on currently available quantum hardware.

\bibliography{apssamp}

\section*{Acknowledgements}

P.T, K.B and A.M are supported by the Polish National Science
Center from funds awarded through the Maestro Grant 
No. DEC-2019/34/A/ST2/00081. 
F.N. is supported in part by: the Japan Science and Technology
Agency (JST) [via the CREST Quantum Frontiers program Grant No.
JPMJCR24I2, the Quantum Leap Flagship Program (Q-LEAP), and the
Moonshot R\&D Grant Number JPMJMS2061], and the Office of Naval
Research (ONR) Global (via Grant No. N62909-23-1-2074).
Our experiments were conducted on the IBM Quantum Network through
the IBM Quantum Hub operated by the Poznań Supercomputing and
Networking Center (PSNC).

\section*{Data and code availability}
All relevant data and code supporting the document is available upon request. Please refer to Patrycja Tulewicz at patrycja.tulewicz@amu.edu.pl.

\section*{Author contributions statement}
K.B. conceived the project, supervised the work and developed MLE for multicopy experiments. P.T. designed and performed the experiments, numerical calculations, and developed the ANN model. P.T and K.B performed analysis of the results and wrote the manuscript.
All authors contributed to the discussions and interpretations of the results and reviewed the manuscript. 

\section*{Competing interests} 
The authors declare no competing interests.
 
\renewcommand{\thefigure}{S\arabic{figure}}
\renewcommand{\thetable}{S\arabic{table}}
\setcounter{figure}{0}
\setcounter{table}{0}

\section*{Appendices}

\section*{Quantum Circuit Implementation}

Accurate construction and manipulation of quantum circuits is key to the successful implementation of our multicopy measurement approach. By carefully manipulating multiple copies of the quantum state, we
have created customized circuits that exactly reproduce
Hong-Ou-Mandel interference.
\begin{figure*}[htb!]
\includegraphics[width=0.9\linewidth]{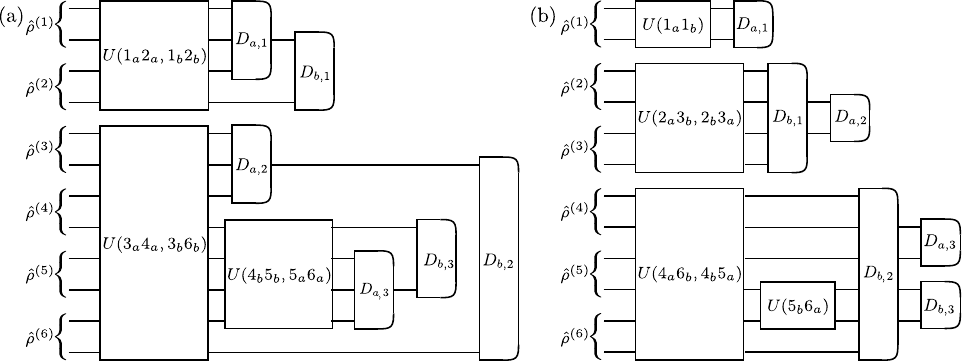}
\caption{\label{fig:singlet} Implementation of singlet projection measurements. (a) Circuit configuration for measuring 
$l_1$, $l_2$, $c_1$, $c_2$, $c_3$, $\bar{c}_3$, $c_4$, and $c_5$. (b) Circuit configuration for measuring $l_0$, 
$\bar{c}_1$, $\bar{c}_2$, $\bar{l}_1$, and $\bar{l}_2$. Gates $U(k_ak_b)$ and $U(k_al_b|l_ak_b)$ are detailed 
in Fig.~\ref{fig:circ}. Detector pairs $D_{a,n}$ and $D_{b,n}$ ($n = 1, 2, 3$) measure coalescence and anticoalescence, 
with total counts $s = a + c$ proportional to the incident pair rate.}
\end{figure*}
\begin{figure*}[htb!]
\includegraphics[width=1\linewidth]{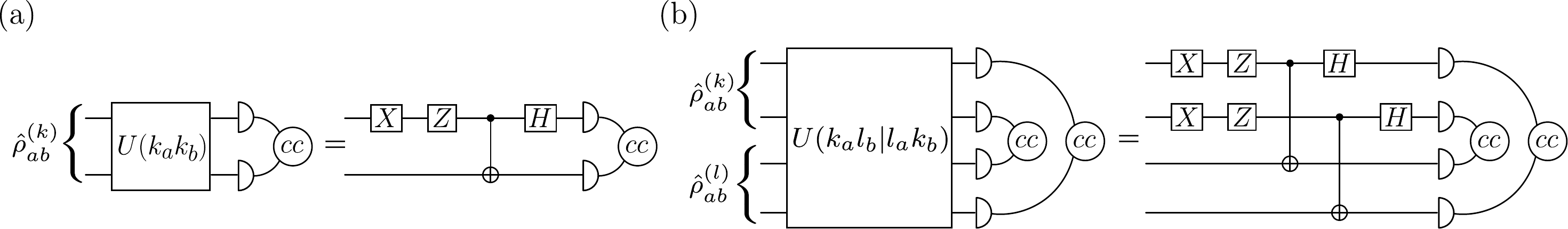}
\caption{\label{fig:circ} Fundamental circuit blocks for singlet projections. (a) Gate $U(k_ak_b)$ operation acting 
on the $k$th entangled qubit pair. (b) Gate $U(k_al_b|l_ak_b)$ acting on two copies of qubit pairs from 
subsystems $a$ and $b$. These blocks form the building elements for all projection measurements.}
\end{figure*}

\subsection*{State--copy Preparation Protocol}

Another important element of our approach is the simultaneous preparation of a large number of identical copies of the quantum state.
We have achieved this in practice through a systematic averaging and qubit mapping process:

\begin{equation}
\hat{\rho}_{\text{final}} = \sum_{i=1}^{N_{\text{maps}}} w_i \hat{\rho}_i,
\end{equation}
where $w_i$ represents optimized weights determined by hardware characteristics
\begin{equation}
w_i = \frac{\exp(-\sum_j \epsilon_{ij})}{\sum_k \exp(-\sum_j \epsilon_{kj})},
\end{equation}
where, $\epsilon_{ij}$ represents the error rate for the $j$th operation in mapping $i$. This averaging method 
significantly improves the fidelity of our multicopy measurements.

\subsection*{Measurement Configurations}

Table~\ref{tab:configs} illustrates various detector settings employed to measure 
different correlation terms in Fig.~\ref{fig:singlet}. The rows correspond to various 
combinations of coalescence ($a$) and summed signal ($s$) over detector pairs 
$D_{a,n}$ and $D_{b,n}$. By projecting these results onto $l_i$ and $c_i$, or their barred 
counterparts, we can efficiently acquire properties such as singlet projections and correlation 
coefficients. This systematic labeling supports repeatability and 
transparency in interpreting the measurement results.
\begin{table}[htb!]
\caption{\label{tab:configs} Detector configurations and their corresponding measurements for the circuits in 
Fig.~\ref{fig:singlet}. Here $a$ represents anticoalescence, $s$ represents the sum signal.}
\begin{center}

\begin{tabular}{cccccc|cc}
\hline\hline
$D_{a,1}$ & $D_{a,2}$ & $D_{a,3}$ & $D_{b,1}$ & $D_{b,2}$ & $D_{b,3}$ & (a) & (b) \\
\hline
$a$ & $s$ & $s$ & $a$ & $s$ & $s$ & $l_1$ & -- \\
$s$ & $a$ & $a$ & $s$ & $a$ & $a$ & $l_2$ & -- \\
$s$ & $s$ & $s$ & $s$ & $s$ & $a$ & $c_1$ & -- \\
$s$ & $s$ & $a$ & $s$ & $s$ & $s$ & $c_2$ & -- \\
$s$ & $a$ & $s$ & $s$ & $a$ & $s$ & $c_3$ & -- \\
$s$ & $s$ & $a$ & $s$ & $a$ & $a$ & $\bar{c}_3$ & -- \\
$s$ & $s$ & $s$ & $a$ & $a$ & $a$ & $c_4$ & -- \\
$s$ & $a$ & $s$ & $s$ & $a$ & $a$ & $c_5$ & -- \\
$a$ & $s$ & $s$ & $s$ & $s$ & $s$ & -- & $l_0$ \\
$s$ & $s$ & $s$ & $a$ & $s$ & $s$ & -- & $\bar{c}_1$ \\
$s$ & $s$ & $s$ & $s$ & $a$ & $a$ & -- & $\bar{c}_2$ \\
$s$ & $a$ & $s$ & $a$ & $s$ & $s$ & -- & $\bar{l}_1$ \\
$s$ & $s$ & $a$ & $s$ & $a$ & $a$ & -- & $\bar{l}_2$ \\
\hline\hline
\end{tabular}
\end{center}

\end{table}

\subsection*{Shot Noise Dependencies}
\begin{figure*}[htb!]
\includegraphics[width=1\linewidth]{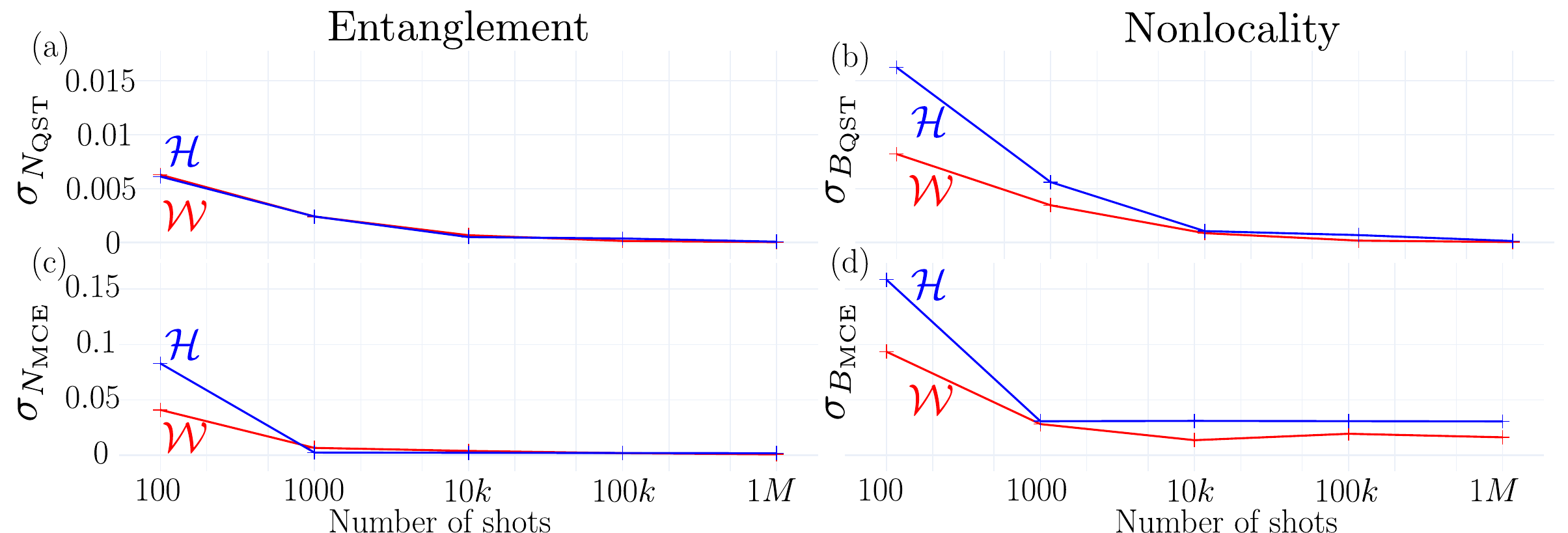}
\caption{\label{fig:shots} Measurement precision analysis. Standard deviations of (a,c) the negativity $N$ and (b,d) 
nonlocality parameter $B$ measurements as functions of shot count. Results shown for both (a,b) QST and (c,d) MCE methods, 
applied to the Werner ($ \mathcal{W}$) and Horodecki ($\mathcal{H}$) states.}
\end{figure*}

Figure~\ref{fig:shots} shows the standard deviations of the negativity $N$ and the 
nonlocality measure $B$ as a function of the number of shots. Panels (a,b) compare QST results 
for the Werner and Horodecki states, respectively, and (c,d) compare the corresponding for MCE. The 
higher the number of shots, the lower the statistical fluctuations; however the rate of 
this improvement depends on the measurement protocol. These plots show that MCE can 
potentially achieve similar or even improved accuracy with fewer shots under certain 
regimes because it was a smaller set of more informative measurements.

\subsection*{Hardware Characteristics}
\begin{figure*}[htb!]
\includegraphics[width=1\linewidth]{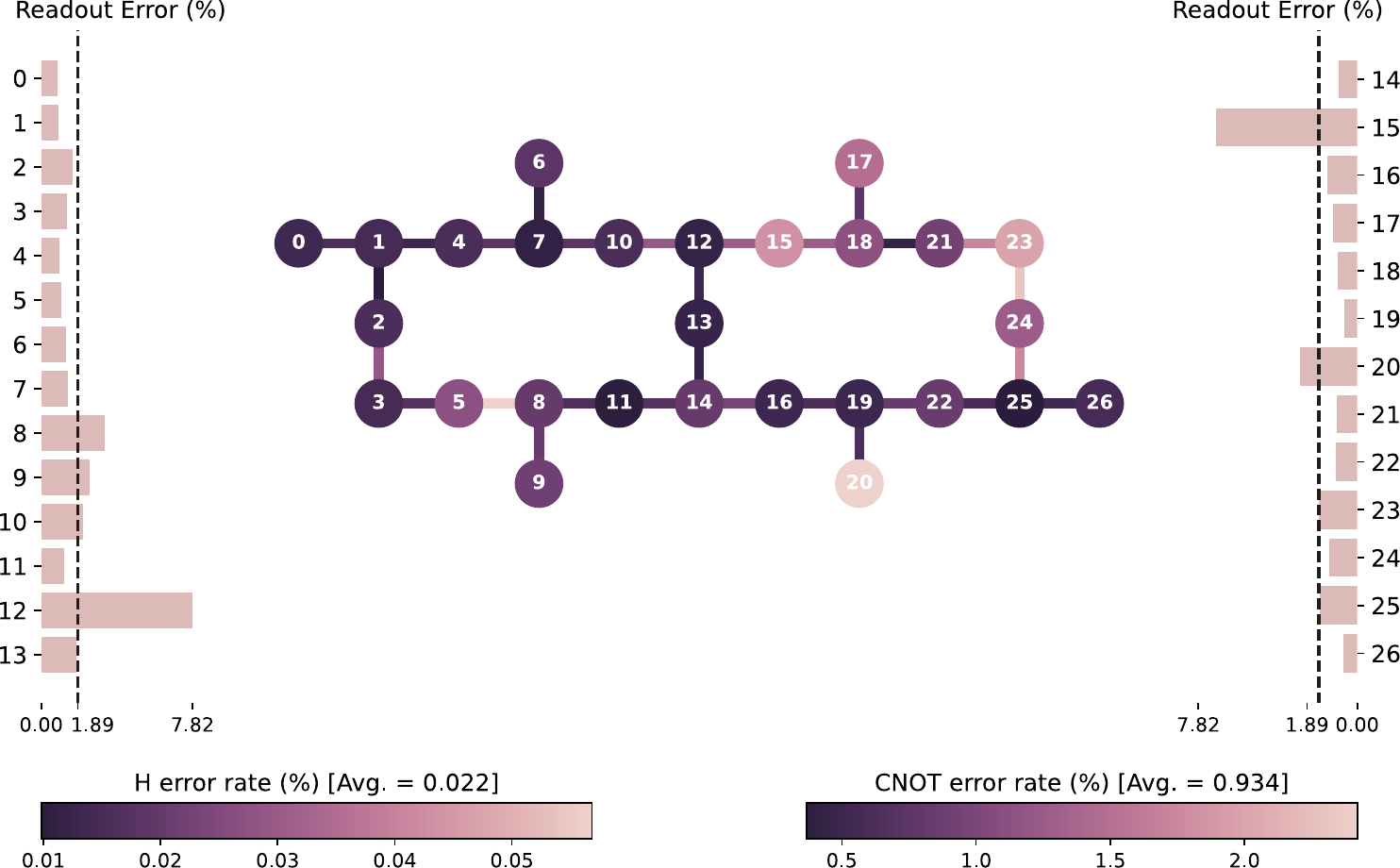}
\caption{\label{fig:processor} Error characterization of the \textit{ibm\_hanoi} processor. Calibration data shown for the 
relevant qubit subset $\{1-5, 7, 10, 12-14, 16, 19\}$ used in our experiments. Here $H$ is the Hadamard gate.}
\end{figure*}

Finally, Fig.~\ref{fig:processor} provides calibration data for all qubits of the \textit{ibm\_hanoi} processor used in our experiment. We show one- and two-qubit gate error histograms and readout infidelities for the subset of qubits $\{1-5, 7, 10, 12-14, 16, 19\}$. The hardware features must be interpreted and adapted so that they can be addressed and optimized using multicopy measurement
circuits to achieve the highest possible fidelity.

\end{document}